\documentclass[
preprint,
prd,
aps,
eqsecnum,
amsmath,
amssymb,
]{revtex4}

\def\slash#1{\, /\kern-0.6em{#1}}

\def\slashchar#1{\setbox0=\hbox{$#1$}           
   \dimen0=\wd0                                 
   \setbox1=\hbox{/} \dimen1=\wd1               
   \ifdim\dimen0>\dimen1                        
      \rlap{\hbox to \dimen0{\hfil/\hfil}}      
      #1                                        
   \else                                        
      \rlap{\hbox to \dimen1{\hfil$#1$\hfil}}   
      /                                         
   \fi}                                         %

\usepackage{graphicx}
\usepackage{color}
\usepackage{float}
\usepackage{subfloat}
\usepackage{subfigure}

\usepackage{hyperref}

\begin{document}

\title{Anomalous chromomagnetic moment of quarks} 
\author{ Ishita Dutta Choudhury}\email{ishitadutta.choudhury@bose.res.in}
\author{Amitabha Lahiri}\email{amitabha@bose.res.in}
\affiliation{S. N. Bose National Centre For Basic Sciences,\\
 Block JD, Sector III, Salt Lake, Kolkata 700098, INDIA}
\date{\today}

\begin{abstract}
We do a perturbative calculation of the anomalous chromomagnetic 
dipole moment of quarks at one-loop, also considering the
effect of a small gauge-invariant mass of the gluon. We find 
partial agreement with a previous calculation, as well as a 
divergence. We explain these results by noting that perturbation
theory is not valid at the energy scales where these calculations
were done, and proceed to give the results at the $M_Z$ scale. 
We find significant variation, of the anomalous moment of the 
light quarks, as a function of gluon mass.

\end{abstract}

\maketitle
\section{Introduction}
The Linear Hadron Collider (LHC), with its recent discovery of
a  125~GeV Higgs boson~\cite{Aad:2012tfa, Chatrchyan:2012ufa}, 
has completed the observation of all fundamental particles 
of the Standard Model. But many questions remain unanswered
about the properties of the low energy particle universe.
The origin of neutrino mass and mechanism of family symmetry 
breaking are the ones that get the most attention, along with
that of whether there are more particles to be found 
as the energy and luminosity of the LHC increases. We are 
interested in another question, one that concerns quantum 
chromodynamics (QCD). The LHC is a QCD machine, and it is 
likely to provide a unique window to precision QCD.

The anomalous magnetic dipole moments of the electron and the 
muon are among the most precisely computed and measured 
quantities in quantum electrodynamics. The corresponding 
quantity for QCD, the anomalous chromomagnetic dipole moment
(CMDM) of quarks, is not so precisely known. Starting from 
a general effective Lagrangian containing the anomalous 
couplings~\cite{Buchmuller:1985jz, Arzt:1994gp}, one may 
look for the contribution of new physics in various processes
by analyzing available data. While some bounds have been 
obtained this way for the new physics contribution to the 
CMDM of the top quark~\cite{Martinez:1996cy, Martinez:2001qs, 
Hioki:2009hm, 
Hioki:2010zu, Kamenik:2011dk, Biswal:2012dr, Choudhury:2012np, 
Labun:2012ra, Degrande:2012gr, Ayazi:2013cba, Hioki:2013hva, 
Hesari:2014hva}, 
this particular quantity has received only sparse attention 
from the community so far. Indeed, it was only very recently 
that an experimental collaboration did an analysis of the 
top-quark CMDM for the first time~\cite{CMS:2014bea}. 
Even less attention seems to have been paid to the 
anomalous CMDM of the other quarks, although some 
calculations have been done for light quarks in the 
context of computing the contribution of quark CMDM 
to nucleon anomalous magnetic moments and electromagnetic 
form factors~\cite{Chang:2010hb}. As the LHC starts 
measuring QCD quantities more accurately, we can expect
more interest in anomalous chromomagnetic dipole moments,
and more generally form factors, of quarks.

Another quantity that is likely to be known more precisely 
through measurements at the LHC is the mass of the gluon. 
There is no Higgs mechanism for QCD as the 
color symmetry of the theory is 
unbroken~\cite{Beringer:1900zz}. On the other hand, a
Proca mass term for the gluon breaks gauge symmetry,
leading to a breakdown of renormalizability as well 
as violation of unitarity at high energy by certain 
tree level amplitudes. However, there are some 
ways a gluon can be massive in a gauge-invariant manner. 

One is due to Cornwall~\cite{Cornwall:1981zr}, who 
suggested that non-zero gluon mass can be
generated dynamically in a theory in which the color 
symmetry remains unbroken.  A dynamically 
generated gluon mass depends on momentum; it must 
also vanish at large momentum so as to maintain
renormalizability of the theory. Another model of a 
massive gluon is the Curci-Ferrari model~\cite{Curci:1976bt}, 
which has a Proca mass term as well as a `gauge-fixing' 
term, uses a quartic ghost interaction to make it 
renormalizable (but not unitary)~\cite{deBoer:1995dh}.
Another is the topological mass generation 
mechanism~\cite{Allen:1991gb}, 
in which an antisymmetric tensor provides a mass to 
the gauge boson via a derivative coupling without
breaking global or gauge symmetries. This model also
appears to be unitary and 
renormalizable~\cite{Hwang:1997er, Lahiri:1997dm, 
Lahiri:2011ic, Lahiri:2001uc}.

Early analyses led to estimates of the gluon mass over a
large range, from $500\pm 200$~MeV~\cite{Cornwall:1981zr}
using numerical calculations of the mass gap, to 
$\simeq 800$~MeV~\cite{Parisi:1980jy} on the basis of 
strong suppression of the end point of the
photon spectrum in radiative $J/ \psi$ decays, to 
$\simeq 1$~GeV~\cite{Field:2001iu} based on analysis
of photon spectra in the processes $J/\psi \to \gamma X$
and $\Upsilon \to \gamma X\,.$ More recently, an analysis
of data from free quark searches found a nominal upper 
limit of ${\cal O}(1)$ MeV~\cite{Nussinov:2010jg}. It is 
clear from these studies that the question of whether 
the gluon has a mass is yet to be settled experimentally.
We can expect that precision measurements of QCD quantities,
such as the anomalous CMDM, will lead to setting bounds 
on gluon mass.

In this paper we calculate the anomalous CMDM of quarks 
at one-loop order, assuming a small mass ($<10$ MeV) 
for the gluon. 
Let us first describe what we plan to calculate, and 
the notation and conventions that we will use. The piece 
of the Lagrangian which governs the quark-gluon coupling 
for a non-zero value of the anomalous CMDM is given by
\begin{equation}
 {\cal L}_{CMDM} = g_{s}\bar{\psi}T _{a}  iF_{2}(q^{2})
 \sigma_{\mu\nu} q^{\nu} \psi\, G^{\mu a}\,,
\label{intro.lag}
\end{equation}
where $\psi$ denotes the quark and $G^{\mu a}$ denotes 
the gluon, 
$g_{s}$ is the strong coupling constant, $T_{a}$ are
the usual SU(3)$_{c}$ generators, and $m$ is the quark 
mass. We will call $F_{2}(q^{2})$ the CMDM, corresponding 
to a momentum transfer $q$. For future reference, we mention
here that in~\cite{Martinez:2001qs, Martinez:2007qf}, 
which calculated 
the CMDM of the top quark, a quantity $\Delta\kappa$ 
was defined by
\begin{equation}
\frac{\Delta\kappa}{4m} = F_2(q^2 = 0)
\end{equation}
and referred to as the CMDM of the quark. 

\begin{figure}[htbp]
\subfigure[]{\label{fig.qed}
\includegraphics[width=0.34\columnwidth]{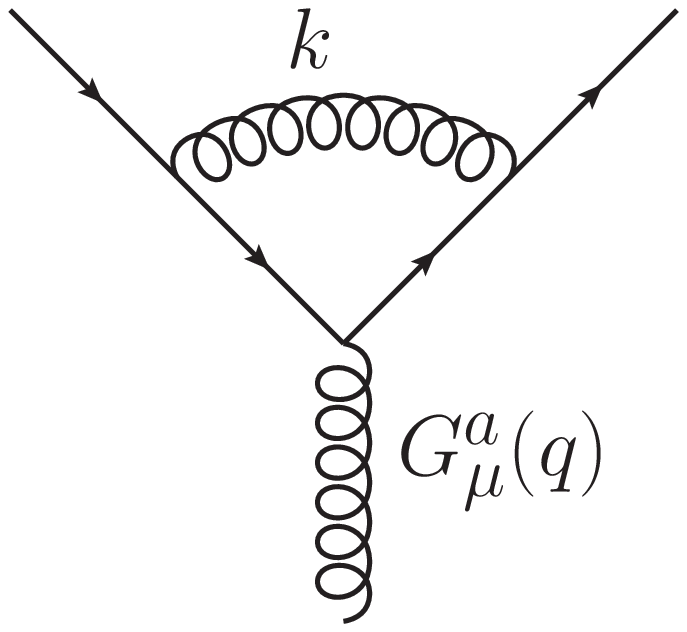}} 
\hspace{1cm}               
\subfigure[]{\label{fig.3g}
\includegraphics[width=0.3\columnwidth]{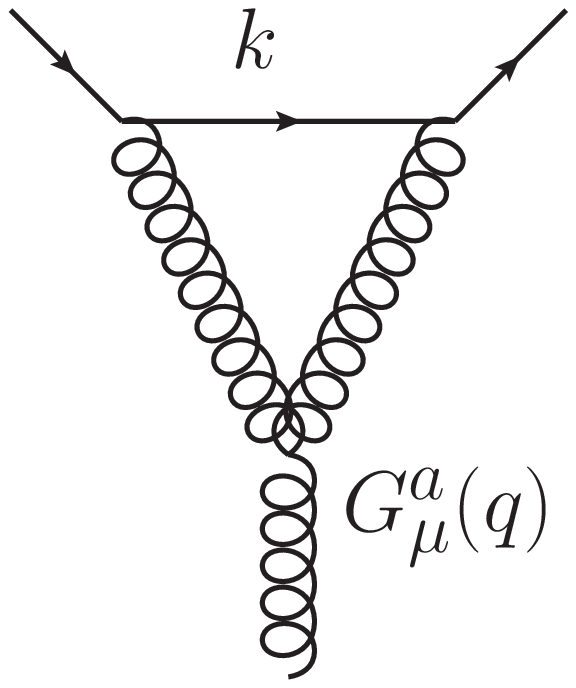}}
\caption{Strong contributions to the anomalous chromomagnetic
  dipole moment of a quark: (a) QED-like diagram; (b) purely 
  non-Abelian contribution.} 
  \label{fig.strong}
\end{figure}

The form factor $F_{2}(q^{2})$ receives contributions 
from both strong and electroweak processes. 
The lowest order QCD contribution comes from two different 
Feynman diagrams, shown in Fig.~\ref{fig.strong}. For the 
diagram in Fig.~\ref{fig.qed} the external gluon directly 
couples to the fermion line in the loop, similar to the 
analogous process of quantum electrodynamics. 
The other diagram, shown in Fig.~\ref{fig.3g}, is purely
non-Abelian in nature, with the external gluon coupling to 
internal gluons. 

\begin{figure}[htbp]
\subfigure[]{\label{fig.eweak}
\includegraphics[width=0.3\columnwidth]{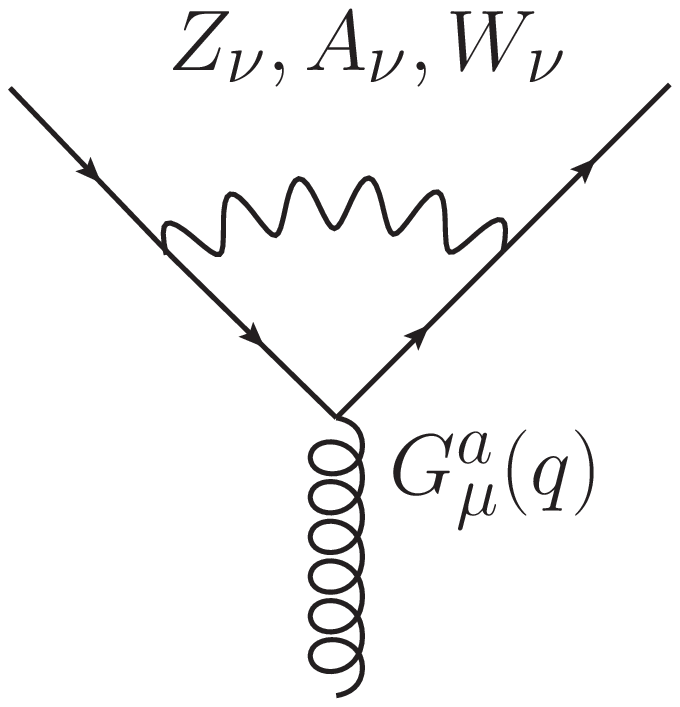}} 
\hspace{1cm}               
\subfigure[]{\label{fig.higgs}
\includegraphics[width=0.3\columnwidth]{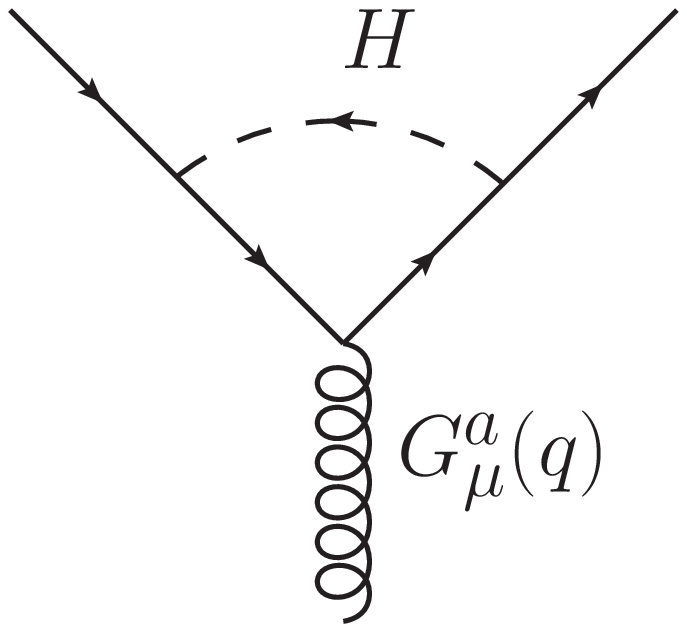}}
\caption{Electroweak contributions to the anomalous chromomagnetic
  dipole moment of a quark: (a) gauge boson exchange;
  (b) Higgs boson exchange.} 
  \label{fig.weak}
\end{figure}

The electroweak contribution to the CMDM comes from loops 
containing the exchange of electroweak gauge bosons 
$\gamma, Z, W$,  or the Higgs boson. 
Using the observed value of the Higgs boson mass,
$m_H = 125.9$ GeV~\cite{Beringer:1900zz} 
completely fixes the contribution from the 
electroweak sector. The relevant vertex corrections 
at the lowest order come from 
the diagrams shown in Fig.~\ref{fig.weak}. 

To begin with, we focus our attention on $F_{2}(0)$, 
calculating this quantity for all quarks and for 
varying gluon mass between 0 and 10 MeV. An analytical 
calculation of the anomalous CMDM was done 
in~\cite{Martinez:2001qs} and later in~\cite{Martinez:2007qf}. 
Also see~\cite{Davydychev:2000rt} for a different method 
of calculation. Apart from the
use of a massive gluon propagator, there is another 
significant way by which our calculations and results 
differ from those in these papers. The results and 
calculations in these papers were done for zero momentum
exchange, i.e. $q^2 = 0\,,$ whereas we do the 
calculations for a general $q^2\,.$ For $q^2 = 0$ the 
diagram of Fig.~\ref{fig.3g} produces a divergent 
contribution to the CMDM for vanishing gluon mass. 
While it is possible to get rid of this infrared 
divergence if we work with zero gluon mass from 
the beginning, this diagram cannot be subtracted out
if the physical gluon has a mass. Then this diagram
contributes to the total, and $F_2(0)$ diverges 
as the gluon mass is taken to zero. We will argue 
below that perturbative $F_{2}(0)$ is not a physically 
sensible quantity anyway, and what we should be 
interested in 
is $F_2(q^2)$ for large values of $-q^2$\,.

We note here that while a calculation for the anomalous 
CMDM was given in~\cite{Martinez:2007qf}, those calculations 
appear to be incorrect. Specifically, the numerical results 
given in~\cite{Martinez:2007qf} are in severe disagreement 
with the calculations given in the appendix of the same 
paper, one being finite and the other divergent. The 
results and calculations are both in disagreement
with ours. Our results for $F_2$ at $q^2 = 0$ appear to 
agree with those in~\cite{Davydychev:2000rt}, whose result
for Fig.~\ref{fig.3g} also diverges at $q^2 =0$ but vanishes
for vanishing quark mass, as does ours. Our formula
for $Z$-exchange agrees with a similar formula 
in~\cite{Fujikawa:1972fe}. 

In Sec.~\ref{calc} we provide an outline of the 
calculation of various contributions to the CMDM 
for $q^2 =0$\,. We then discuss the shortcomings of 
these results and in Sec.~\ref{MZ2} provide the
results for $q^2 = - M_Z^2$\,.

\section{Calculations}\label{calc}
The calculations proceed in a straightforward manner. We 
display only the main steps for the calculation of the 
diagrams in Figs.~\ref{fig.strong} and \ref{fig.weak}. 
We will calculate the form factor $F_2$ at $q^2 = 0\,,$ and
we will take the gluon propagator to be 
\begin{equation}
\Delta_{\mu\nu}^{ab} = -i\delta^{ab} \frac{g_{\mu\nu}}
{k^2 - M^2 + i\epsilon}\,,
\end{equation}
$M$ being the mass of the gluon. There will also be a 
term proportional to $k^\mu k^\nu$ in the propagator, 
the actual form of the term depending on the theory 
that provides a mass to the gluon. But in the diagrams 
of Fig.~\ref{fig.strong}, each internal gluon line 
couples to a conserved $U(1)$ current at least at 
one end. So this term will not contribute 
in these diagrams. The calculations below are relevant for 
of the dynamical mass generation model and the 
Curci-Ferrari model; the topological mass generation 
model has additional diagrams and will be considered
elsewhere.

\subsection{Strong contribution}
The contribution of the diagram in Fig.~\ref{fig.qed} to the vertex
function $\Gamma_{\mu}$ can be written as  
\begin{align}
\Gamma_{\mu}^{(1a)}(q)T^{a}_{ji} &=
-\frac{i}{6}g_s^{2}T^{a}_{ji}\int\dfrac{d^{4}k}{(2\pi)^{4}}
\frac{\gamma_\lambda
(\slashchar{p}'+\slashchar{k}+m)\gamma_\mu(\slashchar{p} +
\slashchar{k}+m)\gamma^\lambda}{[(p'+k)^{2}-m^{2}]
[(p+k)^2-m^2](k^2-M^2)}\,,   
\label{gamma1a}
\end{align}
where we have used $ T^{b}_{jk}T^{a}_{ki'}T^{b}_{i'i}
=-\frac{1}{6}T^{a}_{ji}$.  A straightforward calculation 
gives us the coefficient of 
$i\sigma_{\mu\nu}q^{\nu}\,,$  which is the contribution 
of Fig.~\ref{fig.qed} to the form  factor $F_2(q^2)$\,,
\begin{align}
F^{(1a)}_{2}(q^{2}) &= -\dfrac{8i\alpha_{s}\pi m}{3}
\int \dfrac{d^{4}k}{(2\pi)^{4}}\int\limits_{0}^{1}d\zeta_{1}
\int\limits_{0}^{1}d\zeta_{2}\int\limits_{0}^{1}d\zeta_{3}\,
\delta(1-\zeta_{1} - \zeta_{2} -\zeta_{3})
\nonumber\\ &\qquad\qquad \times 
\dfrac{(\zeta_{1} + \zeta_{2})
(1-\zeta_{1} - \zeta_{2})}
{[k^{2} - (\zeta_{1} + \zeta_{2})^{2} m^{2} 
+ \zeta_{1}\zeta_{2}q^{2} - \zeta_{3}M^{2}]^{3}}\,.
\label{F1aq^2}
\end{align}
For $q^2 = 0$\,, we get 
\begin{equation}
F^{(1a)}_{2}(0)
= -\dfrac{\alpha_{s}}{12m\pi }\int\limits_{0}^{1}d\zeta\dfrac{(1
  - \zeta)^2\zeta}{(1 - \zeta)^{2} +
  \zeta\lambda^{2}}\,,  
\label{F1a0}
\end{equation}
where we have defined 
\begin{equation}
\lambda = \frac{M}{m}\,.
\end{equation}
The result of this integral is included in the plot
of Fig.~\ref{Fig.1a_10mev}, using $\alpha_s(M_Z)\,.$

The value of $F_2(0)$ for the top quark was calculated 
in~\cite{Martinez:2001qs} for vanishing gluon mass, 
using only the diagram of Fig.~\ref{fig.qed}. Our 
result agrees with that value. The calculation for
this diagram, as given in Eq.~(A5) 
of~\cite{Martinez:2007qf}, gives a divergent 
result and is in disagreement with our calculations 
above.

We now calculate the other part of the strong contribution 
to the anomalous chromomagnetic dipole moment coming from 
Fig.~\ref{fig.3g}. The one loop contribution to the vertex 
function $\Gamma_\mu$ for this diagram is given by
\begin{align}
\Gamma_{\mu}^{(1b)} = -i\frac{g_s^2}{4} \int&\frac{d^{4}k}
{(2\pi)^{4}}\dfrac{\gamma^{\lambda}(\slashchar{k}+m)
  \gamma^{\nu}}
{(k^{2}-m^{2})[(k-p)^{2} - M^{2}][(k-p')^{2} - M^{2}]} 
\nonumber\\   & \qquad \times 
[(-2k + p + p')_{\mu}g_{\nu\lambda} + 
(k + p - 2p')_{\nu}g_{\lambda\mu} 
+ (k + p' - 2p)_{\lambda}g_{\mu\nu}]\,,
\label{gamma2}
\end{align}
where we have used  
\begin{equation}
T^{c}_{ji'}T^{b}_{i'i}f_{abc} 
 = -\frac{i}{4}T^{a}_{ji}\,.
\end{equation}
Again after a bit of algebra, we get the
contribution from Fig.~\ref{fig.3g} to the form factor 
$F_{2}(q^{2})$\,,  
\begin{align}
F_{2}^{(1b)} (q^{2}) &= i4\pi m\alpha_{s}
\int\frac{d^{4}k}{(2\pi)^{4}} \int\limits_{0}^{1}d\zeta_{1}
\int\limits_{0}^{1}d\zeta_{2} \int\limits_{0}^{1}d\zeta_{3}\,
\delta(1-\zeta_{1}-\zeta_{2}-\zeta_{3})\nonumber\\  
&\qquad\qquad\times \dfrac{(1-\zeta_{1} -\zeta_{2})(\zeta_{1} + \zeta_{2})}
{[ k^{2} - (\zeta_{1} +\zeta_{2})^{2} m^{2} - (\zeta_{1}
  +\zeta_{2})M^{2} + (\zeta_{1} +\zeta_{2} - \zeta_{3})m^{2} +
  \zeta_{1}\zeta_{2} q^{2}]^{3}}\,, 
\label{F1bq^2}
\end{align}
which for $q^2=0$ gives
\begin{align}
F_{2}^{(1b)} (0) 
&= \dfrac{ \alpha_{s}}{8m\pi }\int\limits_{0}^{1}d\zeta\,
\dfrac{\zeta (1-\zeta)^{2}}{\zeta^{2} 
+ (1 -\zeta)\lambda^{2}} \,.
\label{F1b0}
\end{align}
It was claimed in~\cite{Martinez:2001qs} that this contribution 
to the anomalous chromomagnetic dipole moment of the top
quark vanished, for zero gluon mass. However, here we see
that the integral diverges for $\lambda = 0$\,. We note that 
the formula (A12) in~\cite{Martinez:2007qf}, corresponding to
this diagram, also diverges. A divergence is also found 
in~\cite{Davydychev:2000rt}, vanishing for zero quark mass,
as happens for Eq.~(\ref{F1bq^2}).

We will come back to this point about the divergence later. 
Now we add the results of 
Eq.~(\ref{F1a0}) and Eq.~(\ref{F1b0}), and find the total 
contribution 
of strong interactions to the anomalous CMDM of a quark of 
mass $ m $ at one loop, 
\begin{align}
F_{2}^s (0)&= -\dfrac{\alpha_{s}}{24m\pi}
\int\limits_{0}^{1}d\zeta\,
\frac{\zeta(1 - \zeta)(2 - 5\zeta)}
{(1 - \zeta)^{2} + \zeta\lambda^{2}} \,.
\label{F20s}
\end{align}

\subsection{Electroweak contribution}

Next we calculate the one loop contributions to the anomalous CMDM 
of a quark when the internal line is an electroweak gauge boson or
a Higgs boson. The corresponding diagrams are given in
Fig.~\ref{fig.weak}. 

Let us first consider Fig.~\ref{fig.eweak}, for the 
case where the internal line is a $Z$ boson and the quark 
is an up-type quark. The contribution of this diagram 
to the vertex function $\gamma_{\mu}$ is
\begin{align}
\Gamma^{Z}_{\mu}
& =   \dfrac{2 i e^{2}}{\cos^2\theta
  \sin^2\theta} \int\dfrac{d^{4}k}{(2\pi)^{4}}
  \int\limits_{0}^{1} 
d\zeta_{1}\int\limits_{0}^{1}d\zeta_{2}
\int\limits_{0}^{1}d\zeta_{3}  \,
\dfrac{N^{Z}_{\mu}(k)
\delta(1-\zeta_{1}-\zeta_{2}-\zeta_{3})}{D_{Z}^{3}}\,,
\label{gamma.Zu}
\end{align}
where we have written
\begin{equation}
D_{Z} = k^{2} + 2k(\zeta_{1}p' + \zeta_{2}p) -
\zeta_{3}M^{2}_{Z}\,, 
\label{D3k}
\end{equation}
and
\begin{equation}
N^{Z}_{\mu}(k) = \gamma_{\nu}\left(-\frac{1}{2}L 
+ \frac{2}{3}\sin^{2}\theta\right)
\left(\slashchar{k}+\slashchar{p'}+m_u\right) \gamma_{\mu}
\left(\slashchar{k}+\slashchar{p} + m_u\right)\gamma^{\nu} 
\left(-\frac{1}{2}L + \frac{2}{3}\sin^{2}\theta\right)\,.
\label{numerator_k_z}
\end{equation}
Here $m_u$ is the mass of the up-type quark, $M_Z$ is the mass of
the $Z$ boson, and $\theta\equiv \theta_W$ is the weak mixing angle.
As before, after a little algebra, we can rewrite the integral as
\begin{align}
\Gamma^Z_{\mu} &= \dfrac{2ie^{2}}{(\cos\theta \sin\theta)^{2}}
\int\dfrac{d^{4}k}{(2\pi)^{4}}
\int\limits_{0}^{1}d\zeta_{1}\int\limits_{0}^{1}d\zeta_{2} 
\int\limits_{0}^{1}d\zeta_{3}\,
\delta (1-\zeta_{1} - \zeta_{2} -\zeta_{3})\nonumber\\
&\qquad \qquad \times  \dfrac{N^{Z}_{\mu}(k-\zeta_{1}p' -
  \zeta_{2}p)  
+ N^{Z}_{\mu}(k-\zeta_{1}p - \zeta_{2}p')}
{2\left[ k^{2} -(\zeta_{1}+\zeta_{2})^{2}m^{2} +\zeta_{1}\zeta_{2}
    q^{2} 
- \zeta_{3}M^{2}_{Z}\right]^{3}}\,.
\label{gammaz}
\end{align}

Ignoring all terms that do not contribute to $F_2(q^2)\,,$
such as those proportional to $\gamma_5\,,$ we find 
that the contribution from 
$ N^{Z}_{\mu}(k-\zeta_{1}p'-\zeta_{2}p) $ is
\begin{equation}
m\left[-\frac{1}{2} (1-\zeta_{1}) +(-\frac{4}{3}\sin^{2}\theta 
+\frac{16}{9}\sin^{4}\theta) \zeta_{1}\right](1-\zeta_{1}-\zeta_{2})\,.
\label{numw.a.1}
\end{equation}
Similarly, the contribution from $
N^{Z}_{\mu}(k-\zeta_{2}p'-\zeta_{1}p) $ is 
\begin{equation}
m\left[-\frac{1}{2} (1-\zeta_{2}) +(-\frac{4}{3}\sin^{2}\theta 
+\frac{16}{9}\sin^{4}\theta)
\zeta_{2}\right](1-\zeta_{1}-\zeta_{2})\,. 
\label{numw.a.2}
\end{equation}

Combining Eq.s~(\ref{numw.a.1}), (\ref{numw.a.2}) and
(\ref{gammaz}), we find that the contribution 
to the anomalous CMDM of up-type quarks from
a $Z$-mediated diagram in Fig.~\ref{fig.eweak}  is
given by
\begin{align}
F^{Z}_{2}(q^{2}) &= -4\sqrt 2 i G_F M_Z^2 m_u
\int\dfrac{d^{4}k}{(2\pi)^{4}}
  \int\limits_{0}^{1}d\zeta_{1}
\int\limits_{0}^{1}d\zeta_{2}   
\int\limits_{0}^{1}d\zeta_{3}\,
\delta (1-\zeta_{1} - \zeta_{2} -\zeta_{3}) \nonumber\\
&\quad \times \dfrac{(\zeta_{1} +\zeta_{2}-1)+ \left(\frac{1}{2}  
- \frac{4}{3}\sin^{2}\theta +\frac{16}{9}\sin^{4}\theta\right)
(\zeta_{1} +\zeta_{2})(1-\zeta_{1}-\zeta_{2})}
{[k^{2} - (\zeta_{1}+\zeta_{2})^{2} m^{2}_{u} 
+\zeta_{1}\zeta_{2}q^{2}-\zeta_{3}M^{2}_{Z}]^{3}}\,,\nonumber \\
\label{FZq2}
\end{align}
so that 
\begin{equation}
F^{Z}_{2}(0) = -\frac{G_F M_Z^2}{4\sqrt 2\pi^2 m_u}
 \int\limits_{0}^{1}d\zeta\,\frac{\zeta(1-\zeta)
\left[-1 + \left(\frac{1}{2} - \frac{4}{3}\sin^{2}\theta +
\frac{16}{9}\sin^{4}\theta)(1-\zeta\right)\right]}
{(1-\zeta)^{2} +\zeta\lambda^{2}_{Z, u}}\,,
\label{FZ0u}
\end{equation}
where we have written $\lambda_{Z, u} =
\displaystyle{\frac{M_Z}{m_u}}\,.$ Our result 
Eq.~(\ref{FZq2}) agrees exactly, after
integration over $k\,,$ with Eq.~(A4) 
of~\cite{Fujikawa:1972fe} upon making the substitutions
appropriate to an up-type quark, 
\begin{equation}
a = \frac{2}{3}\frac{g'^2}{(g^2 + g'^2)^{\frac12}}\,, \qquad
b = \frac{g'^2 - 3 g^2}{(g^2 + g'^2)^{\frac12}}\,,
\end{equation}
where $a$ and $b$ are coupling constants used 
in~\cite{Fujikawa:1972fe}.

Similarly, the contribution from a $Z$-mediated diagram in
Fig~\ref{fig.eweak} to the anomalous CMDM of a down type 
quark is
\begin{equation}
F^{Z}_{2}(0) = -\frac{G_F M_Z^2}{4\sqrt 2\pi^2 m_d}
\int\limits_{0}^{1}d\zeta
\dfrac{\zeta(\zeta-1)+ (\frac{1}{2} 
- \frac{2}{3}\sin^{2}\theta +\frac{4}{9}\sin^{4}\theta)
\zeta(1-\zeta)^{2}}{(1-\zeta)^{2} 
+\zeta\lambda^{2}_{Z,d}}\,,
\label{FZ0d}
\end{equation}
with $\lambda_{Z, d} = \displaystyle{\frac{M_Z}{m_d}}\,,$
where $m_d$ is the mass of the down-type quark.

Next we consider the diagram Fig.~\ref{fig.eweak} with 
a photon line in the loop. The calculation for this diagram 
is completely straightforward. For up-type quarks 
we calculate the contribution to be
\begin{equation}
F^{A}_{2}(0)=\frac{\alpha}{9\pi m_u}\,.
\label{FA0u}
\end{equation}
Similarly, the contribution of the photon-mediated diagram 
in  Fig.~\ref{fig.eweak} to the 
anomalous CMDM of a down-type quark is
\begin{equation}
F^{A}_{2}(0)=\frac{ \alpha}{36\pi m_d}\,.
\label{FA0d}
\end{equation}

When a $W$ is in the loop of Fig.~\ref{fig.eweak}, 
the contribution of the diagram to the vertex 
function of an up-type quark is
\begin{align}
\Gamma^{W}_{\mu} & = 
\dfrac{ie^{2}}{2\sin^{2}\theta} \int\dfrac{d^{4}k}{(2\pi)^{4}}
\dfrac{\gamma_{\nu}L(\slashchar{k}+\slashchar{p'}
+m_{d})\gamma_{\mu}(\slashchar{k}+\slashchar{p} 
+ m_{d})\gamma^{\nu}L}{[(k+p')^{2} - m^{2}_{d}]
[(k+p)^{2} - m^{2}_{d}](k^{2} -M^{2}_{W})}\,.
\label{gamma w}
\end{align}
The form factor is easy to calculate,
\begin{equation}
F^{W}_{2}(0) = -\frac{G_F M_W^2}{4\sqrt 2\pi^2 m_d}
\int\limits_{0}^{1}d\zeta\,
\dfrac{(1-\zeta^{2})\zeta}{(1-\zeta)^{2}  +\zeta\lambda^{2}_{W,d}}\,,
\label{FW0u}
\end{equation}
where we have written, similarly to earlier definitions,
$\lambda^{2}_{W,d}=\frac{M^{2}_{W}}{m^{2}_{d}}\,.$

We similarly calculate the contribution of the 
$W$-mediated diagram to the anomalous CMDM of 
a down type quark to be
\begin{equation}
F^W_2(0) = -\frac{G_F M_W^2}{4\sqrt 2\pi^2 m_u}
\int\limits_{0}^{1}d\zeta\,
\dfrac{(1-\zeta^{2})\zeta}{  (1-\zeta)^{2}  + \zeta\lambda^{2}_{W,u}}\,,
\end{equation}
with $\lambda^{2}_{W,u} = \frac{M^{2}_{W}}{m^{2}_{u}}\,.$

Finally, we consider the diagram with the Higgs boson in the loop
of Fig.~\ref{fig.eweak}. The vertex function for a quark 
for this diagram is
\begin{align}
\Gamma^{H}_{\mu}
& =  -\dfrac{ i
  e^{2}}{2\sin^2\theta}\frac{m^{2}}
{M^{2}_{W}}\int\dfrac{d^{4}k}{(2\pi)^{4}}\,
\dfrac{(\slashchar{k}+\slashchar{p'} +m)
\gamma_{\mu}(\slashchar{k}+\slashchar{p} + m)}
{\left((p+k)^2 - m^2 \right)\left((p'+k)^2 - m^2\right)
\left(k^2 - M^2_H\right)}\,.
\label{gammaH}
\end{align}
Here $M_{H}$ is the mass of the Higgs boson. Proceeding as in the 
previous cases, we obtain the contribution to the anomalous CMDM of 
both up-type and down-type quarks,
\begin{equation}
F^{H}_{2}(0) = \frac{G_F m}{4\sqrt 2\pi^2}
\int\limits_{0}^{1}d\zeta
\dfrac{(1+\zeta)(1-\zeta)^{2}}{(1-\zeta)^{2}  +\zeta\lambda^{2}_{H,i}}\,.
\end{equation}
with $\lambda^{2}_{H,i} = \frac{M^{2}_{H}}{m_i^{2}}\,$ 
and $i$ denotes the quark flavor.

In Table~\ref{table:weak.q=0} we display the total 
contribution to $\Delta\kappa = 4m F_2(0)$ from weak 
interactions for each quark. The integrations were done 
using Mathematica~\cite{Wolfram:2012}.
\begin{table} [htbp]
\centering
\begin{tabular}{cccccc}
\hline\hline
Quark & $Z$ & $A$ & $W$ & $H$ & Total \\[0.1 ex]
\hline
u & $ 0$ & $10.32 \times 10^{-4} $& $ 0$ & $ 0$ & $10.32 \times 10^{-4} $\\
d &$0$ & $2.58 \times 10^{-4}$ &$ 0$ & $0$ & $2.58 \times 10^{-4}$\\
c & $ 0$ & $10.32 \times 10^{-4}$ & $ 0$ & $0$ &$10.32 \times 10^{-4}$\\ 
s & $0 $& $2.58 \times 10^{-4}$ & $ 0$ & $ 0 $& $2.58 \times 10^{-4}$\\
t & $26.26\times 10^{-4}$& $10.32 \times 10^{-4}$
 & $-3.93 \times 10^{-4}$ & $154.61 \times 10^{-4}$ & $187.26\times 10^{-4}$\\
b & $ 0$ & $2.58 \times 10^{-4}$ & $-1.01 \times 10^{-4}$ & $ 0$ & $1.57 \times 10^{-4}$ \\ 
[1 ex]
\hline
\end{tabular}
\caption{Weak contribution to the quark anomalous CMDM $\Delta \kappa$}
\label{table:weak.q=0}
\end{table}
For these calculations, we have taken $\alpha = (137.036)^{-1}\,,$ 
the Fermi constant $G_F = 1.16638 \times 10^{-5}$ 
GeV$^{-2}\,,$ and the current quark masses $m_u = 2.3$ MeV, 
$m_d = 4.8$ MeV, $m_s = 95$ MeV, $m_c = 1.275$ GeV, 
$m_b = 4.18$ GeV, as given in~\cite{Beringer:1900zz}. The zeroes
in the table represent numbers smaller that $10^{-7}\,.$

\begin{figure}[htbp]
\includegraphics[width=0.8\columnwidth]{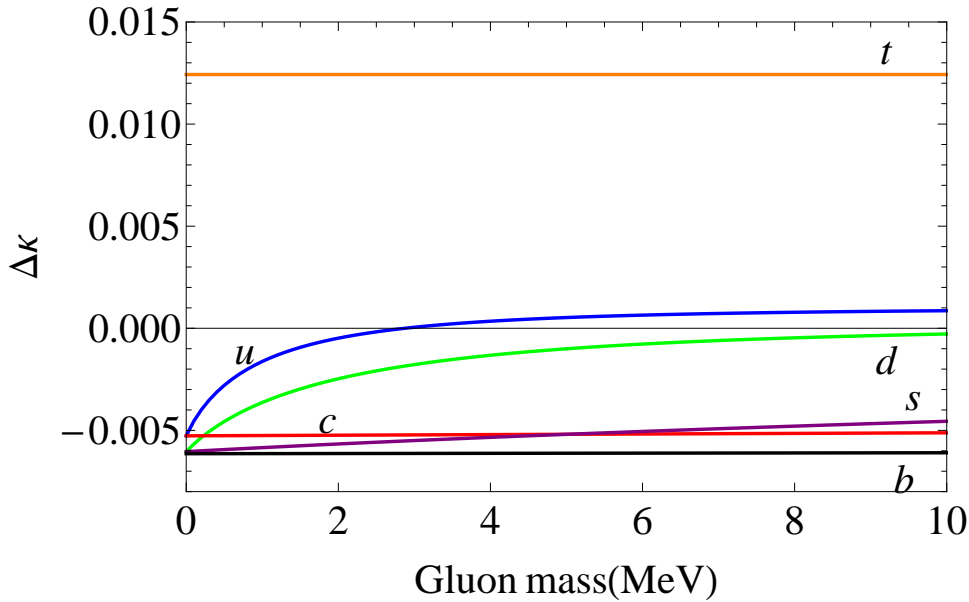} 
\caption{Contribution to the anomalous CMDM from
 weak interactions and Fig.~\ref{fig.qed}.}\ 
\label{Fig.1a_10mev}
\end{figure}

To these numbers we have to add the contributions from the 
strong interactions. The result of adding only
Eq.~(\ref{F1a0}) to the weak contributions 
is plotted as a function of gluon mass in 
Fig.~\ref{Fig.1a_10mev}. This number
does not vary significantly with the mass of the gluons for 
the heavier quarks. That is expected, as the dependence 
on gluon mass $M$ is through the ratio $M/m$, so for large 
enough $m\,,$ small variations in $M$ become negligible.

The contribution from Fig.~\ref{fig.3g} is infrared 
divergent when the gluon is massless. This is because
the three-gluon vertex in Fig.~\ref{fig.3g}, 
with all gluons massless, causes a divergence called 
the mass singularity~\cite{Muta:2010, 
Carazzone:1975cc}, when the external gluon is on-shell. 
If the gluon has a non-zero mass, this diagram 
contributes finitely to the anomalous CMDM
of a quark. As mentioned earlier, if the gluon
is taken to be massless from the beginning, it is
possible to remove this divergence, but if the 
gluon has a mass, this diagram cannot be removed,
and its contribution will diverge when the gluon
mass is taken to zero.

\begin{figure}[htbp]
\centering
\includegraphics[width=0.8\columnwidth]{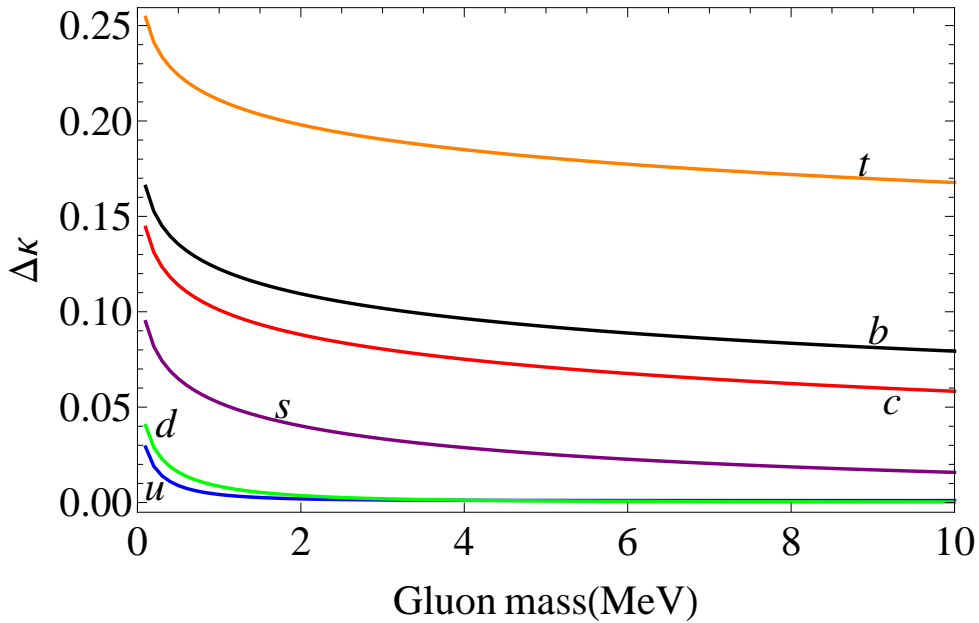}
\caption{Anomalous CMDM of quarks: dependence on gluon mass}\ 
\label{total_10mev}
\end{figure}

We can see this in Fig.~\ref{total_10mev}, where 
we have plotted the total $\Delta\kappa$ for each of 
the quarks, as a function of gluon mass. The contribution
from Fig.~\ref{fig.3g} dominates, and the plots go to 
negative infinity as the gluon mass is taken to zero, but 
rise quickly as the gluon mass is increased. 
%

\section{Anomalous chromomagnetic moment at $q^2 = -M^2_Z$}\label{MZ2}
In the previous section we calculated the anomalous chromomagnetic
dipole moment of each quark by calculating the quark-gluon vertex
form factor $F_2(q^2)$ at $q^2 = 0\,.$ However, this definition is
problematic, since it requires using perturbation theory at 
zero energy, where it is not valid for strong interactions. A 
related issue is that the measured values of the physical 
constants and masses pertaining to strong interactions are 
known at high energies, not at $q^2 = 0\,.$ However, we can use 
the same techniques to calculate the form 
factor $F_2(q^2)$, and thus the anomalous CMDM at 
a higher energy scale. Let us calculate $F_2$ at energy
corresponding to the $Z$-mass, i.e. at $q^2 = - M_Z^2\,.$ 

As before, we calculate $F_2$ by adding up the contributions 
from the diagrams in Fig.s~\ref{fig.strong} and \ref{fig.weak}\,.
Using Eq.~(\ref{F1aq^2}) we can write the contribution from 
the diagram in Fig.~\ref{fig.qed} as
\begin{align}
F^{(1a)}_{2}(q^{2} = -M_Z^{2}) &= -\dfrac{8i\alpha_{s}\pi m}{3}
\int \dfrac{d^{4}k}{(2\pi)^{4}}\int\limits_{0}^{1}d\zeta_{1}
\int\limits_{0}^{1}d\zeta_{2}\int\limits_{0}^{1}d\zeta_{3}\,
\delta(1-\zeta_{1} - \zeta_{2} -\zeta_{3})\nonumber\\
&\qquad\qquad \times \dfrac{(\zeta_{1} + \zeta_{2})
(1-\zeta_{1} - \zeta_{2})}
{[k^{2} - (\zeta_{1} + \zeta_{2})^{2} m^{2} 
- \zeta_{1}\zeta_{2}M_Z^{2} - \zeta_{3}M^{2}]^{3}}\nonumber\\
&=-\dfrac{\alpha_{s}}{12\pi m}\int\limits_{0}^{1}d\zeta_{3}\int
\limits_{0}^{1-\zeta_{3}}d\zeta_{2}\dfrac{(1- \zeta_{3})
\zeta_{3}}
{ (1- \zeta_{3})^{2}  
+ (1-\zeta_{2}-\zeta_{3})\zeta_{2}\lambda_{Z,i}^{2} + 
\zeta_{3}\lambda^{2}}
\label{F1aM_Z^2}
\end{align}
for the quark flavor $i\,,$ with $\lambda$ and $\lambda_{Z,i}$ 
as defined earlier. Similarly, the contribution from 
Fig.~\ref{fig.3g} is given by Eq.~(\ref{F1bq^2}) 
at $q^2 = -M_Z^2\,,$
\begin{align}
F_{2}^{(1b)} (q^{2}= -M_Z^{2}) &= 4i\pi m\alpha_{s}
\int\frac{d^{4}k}{(2\pi)^{4}} \int\limits_{0}^{1}d\zeta_{1}
\int\limits_{0}^{1}d\zeta_{2} \int\limits_{0}^{1}d\zeta_{3}\,
\delta(1-\zeta_{1}-\zeta_{2}-\zeta_{3})\nonumber\\  
&\qquad\times  \dfrac{(1-\zeta_{1} -\zeta_{2})(\zeta_{1} + \zeta_{2})}
{[ k^{2} - (\zeta_{1} +\zeta_{2})^{2} m^{2} - (\zeta_{1}
  +\zeta_{2})M^{2} + (\zeta_{1} +\zeta_{2} - \zeta_{3})m^{2} -
  \zeta_{1}\zeta_{2}M_Z^{2}]^{3}}\nonumber\\
&=\dfrac{\alpha_{s}}{8\pi m}\int\limits_{0}^{1}d\zeta_{3}\int
\limits_{0}^{1-\zeta_{3}}d\zeta_{2}\,\dfrac{(1- \zeta_{3})
\zeta_{3}}
{ \zeta_{3}^{2}  
+(1- \zeta_{2}-\zeta_{3})\zeta_{2}\lambda_{Z,i}^{2} 
+(1- \zeta_{3})\lambda^{2}}\,.
\label{F1bM_Z^2}
\end{align}

Next we consider Fig.~\ref{fig.eweak} for an up-type
quark with a $Z$ boson in the internal line.
Using Eq.~(\ref{FZq2}) we calculate the contribution 
of this diagram to the form factor $F_2(-M_Z^2)$ as
\begin{align}
F^{(Z)}_{2}(q^2 = -M_Z^{2}) &= -\frac{i m_{u} e^2}{\cos^2\theta
  \sin^2\theta}\int\limits_{0}^{1}d\zeta_{1}
\int\limits_{0}^{1}d\zeta_{2}   
\int\limits_{0}^{1}d\zeta_{3}\int\dfrac{d^{4}k}{(2\pi)^{4}}\,
\delta (1-\zeta_{1} - \zeta_{2} -\zeta_{3}) \nonumber\\
&\qquad \times \dfrac{(\zeta_{1} +\zeta_{2}-1)+ \left(\frac{1}{2}  
- \frac{4}{3}\sin^{2}\theta +\frac{16}{9}\sin^{4}\theta\right)
(\zeta_{1} +\zeta_{2})(1-\zeta_{1}-\zeta_{2})}
{[k^{2} - (\zeta_{1}+\zeta_{2})^{2} m^{2}_{u} 
-\zeta_{1}\zeta_{2}M_Z^{2}-\zeta_{3}M^{2}_{Z}]^{3}}\,,\nonumber \\
&=-\dfrac{G_F M_Z^2}{4\sqrt{2}\pi^2 m_u}
\int\limits_{0}^{1}d\zeta_{3}\int
\limits_{0}^{1-\zeta_{3}}d\zeta_{2}\dfrac{-\zeta_{3} +\left(\frac{1}{2}  
- \frac{4}{3}\sin^{2}\theta 
+\frac{16}{9}\sin^{4}\theta\right)\zeta_{3}(1-\zeta_{3})}{(1-\zeta_{3})^{2}+\lambda_{Z,u}\zeta_{2}
(1-\zeta_{2}-\zeta_{3})+\lambda_{Z,u}\zeta_{3}}\,.
\label{FZuM_Z^2}
\end{align}
Similarly, the contribution from Fig.~\ref{fig.eweak} for a down 
type quark with an internal $Z$ boson line is given by
\begin{align}
F^{(Z)}_{2}(q^{2}= -M_Z^{2}) 
&= -\dfrac{G_F M_Z^2}{4\sqrt{2}\pi^2 m_d}
\int\limits_{0}^{1}d\zeta_{3}\int
\limits_{0}^{1-\zeta_{3}}d\zeta_{2}\dfrac{-\zeta_{3} +\left(\frac{1}{2}  
- \frac{2}{3}\sin^{2}\theta +\frac{4}{9}\sin^{4}\theta\right)\zeta_{3}(1-\zeta_{3})}{(1-\zeta_{3})^{2}+\lambda_{Z,d}\zeta_{2}(1-\zeta_{2}-\zeta_{3})
+\lambda_{Z,d}\zeta_{3}}\,.
\label{FZdM_Z^2}
\end{align}
For the diagram in Fig.~\ref{fig.eweak} with photon in the loop, 
the contribution for up-type quarks is 
\begin{align}
 F_{2}^{A}(q^{2}= - M_Z^{2}) &= \frac{2\alpha}{9\pi m_u}
 \int\limits_{0}^{1}d\zeta_{3}\int \limits_{0}^{1-\zeta_{3}} 
 d\zeta_{2} \dfrac{\zeta_{3}(1-\zeta_{3})}{(1-\zeta_{3})^{2}
 + \zeta_{2}(1-\zeta_{2}-\zeta_{3})\lambda_{Z,u}^{2}}\,,
\end{align}
while for the down type quarks it is 
\begin{align}
 F_{2}^{A}(q^{2}= - M_Z^{2}) &= \frac{\alpha}{18\pi m_d}
 \int\limits_{0}^{1}d\zeta_{3}\int \limits_{0}^{1-\zeta_{3}} 
 d\zeta_{2} \dfrac{\zeta_{3}(1-\zeta_{3})}{(1-\zeta_{3})^{2}
 + \zeta_{2}(1-\zeta_{2}-\zeta_{3})\lambda_{Z,d}^{2}}\,.
\end{align}

Next we consider the diagram in Fig.~\ref{fig.eweak} 
with a $W$ boson in the loop. According to Eq.~(\ref{gamma w}) 
the contribution to $ F_{2}(q^{2}= - M_Z^{2})$ for an up 
type quark is 
\begin{align}
F_{2}^{W}(q^{2}= - M_Z^{2}) &=-\dfrac{ i e^2 m_{d}}
{\sin^{2}\theta} \int\limits_{0}^{1}d\zeta_{1}
\int\limits_{0}^{1}d\zeta_{2}   
\int\limits_{0}^{1}d\zeta_{3}\int\dfrac{d^{4}k}{(2\pi)^{4}}\,
\delta (1-\zeta_{1} - \zeta_{2} -\zeta_{3}) \nonumber\\
&\qquad \times \dfrac{(1-\zeta_{1} -\zeta_{2})(2-\zeta_{1} -\zeta_{2})}
{[k^{2}-(\zeta_{1}+\zeta_{2})^{2}-
\zeta_{1}\zeta_{2}\lambda_{Z,d}^{2}-
\zeta_{3}\lambda_{W,d}]^{3}}\,,\nonumber \\
&= -\frac{G_F M_W^2}{4\sqrt{2}\pi^2 m_{d}}
\int\limits_{0}^{1}d\zeta_{3}\int
\limits_{0}^{1-\zeta_{3}} d\zeta_{2}
\dfrac{\zeta_{3}(1-\zeta_{3})}{(1-\zeta_{3})^{2}
+\zeta_{2}(1-\zeta_{2}-\zeta_{3})\lambda_{Z,d}^{2} 
+ \zeta_{3}\lambda_{W,d}^{2}}\,.
\end{align}
Similarly, for the down type quark we obtain
\begin{align}
F_{2}^{W}(q^{2}= - M_Z^{2}) &=
-\frac{G_F M_W^2}{4\sqrt{2}\pi^2 m_{u}}
\int\limits_{0}^{1}d\zeta_{3}\int
\limits_{0}^{1-\zeta_{3}} d\zeta_{2}
\dfrac{\zeta_{3}(1-\zeta_{3})}{(1-\zeta_{3})^{2}
+\zeta_{2}(1-\zeta_{2}-\zeta_{3})\lambda_{Z,u}^{2} 
+ \zeta_{3}\lambda_{W,u}^{2}}\,.
\end{align}

The contribution from the diagram in Fig.~\ref{fig.higgs} to 
$ F_{2}(q^{2}= - M_Z^{2})$ of a quark is obtained from 
Eq.~(\ref{gammaH}) as
\begin{align}
 F_{2}^{H}(q^{2}= - M_Z^{2}) = 
 \dfrac{G_F m_i} {4\sqrt{2}\pi^2}
 \int\limits_{0}^{1}d\zeta_{3}\int \limits_{0}^{1-\zeta_{3}} 
 d\zeta_{2} \dfrac{(1+\zeta_{3})(1-\zeta_{3})}{(1-\zeta_{3})^{2}
 + \zeta_{2}(1-\zeta_{2}-\zeta_{3})\lambda_{Z,i}^{2}
 + \zeta_{3}\lambda_{H,i}^{2}}\,,
\end{align}
where $i$ denotes the quark flavor as before.

\begin{table} [htbp]
\centering
\begin{tabular}{cccccc}
\hline\hline
Quark & $Z$ & $A$ & $W$ & $H$ & Total \\[0.1 ex]
\hline
u & $1.29\times10^{-12}$ & $29.79 \times 10^{-12} $& $-4.41 \times 10^{-12}$ 
& $0$ & $26.67 \times 10^{-12}$\\
d &$5.50 \times 10^{-12}$ & $30.18 \times 10^{-12}$ &
$ -4.41 \times 10^{-12}$ & $ 0$ & $31.27 \times 10^{-12}$\\
c & $3.98 \times 10^{-7}$ & $36.91 \times 10^{-7}$ & $-0.48 \times 10^{-7}$ 
& $0$ & $40.41 \times 10^{-7}$\\ 
s & $0.22 \times 10^{-8} $& $0.82 \times 10^{-8}$ & $-4.82\times 10^{-8}$ 
& $ 0 $& $-3.78 \times 10^{-8}$\\
t & $25.66 \times 10^{-4}$& $10.57 \times 10^{-4}$ & $-2.87 \times 10^{-4}$ 
& $149.91 \times 10^{-4}$ & $183.27 \times 10^{-4}$\\
b & $4.16 \times 10^{-6}$ & $7.14 \times 10^{-6}$ & $-98.72 \times 10^{-6}$ 
& $0.03 \times 10^{-6}$ & $-87.39 \times 10^{-6}$ \\ [1 ex]
\hline
\end{tabular}
\caption{Electroweak contribution to $4m F_{2}(q^{2}= - M_Z^{2})$ 
of each quark}
\label{table:weak-MZ}
\end{table} 

\begin{figure}[htbp]
\subfigure[]{\label{fig.mz_u}
\includegraphics[width=0.3\textwidth]{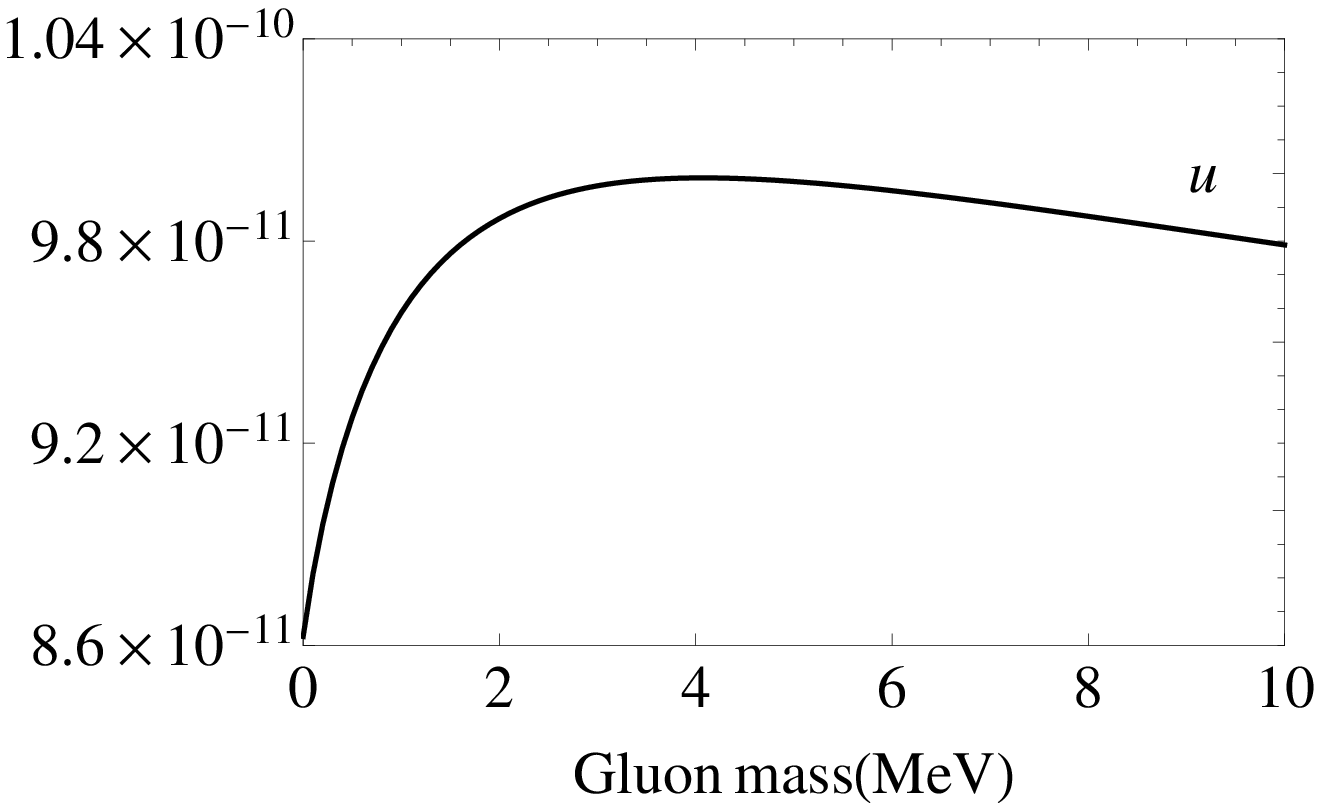}} 
\hspace{0.02\textwidth}               
\subfigure[]{\label{fig.mz_d}
\includegraphics[width=0.3\textwidth]{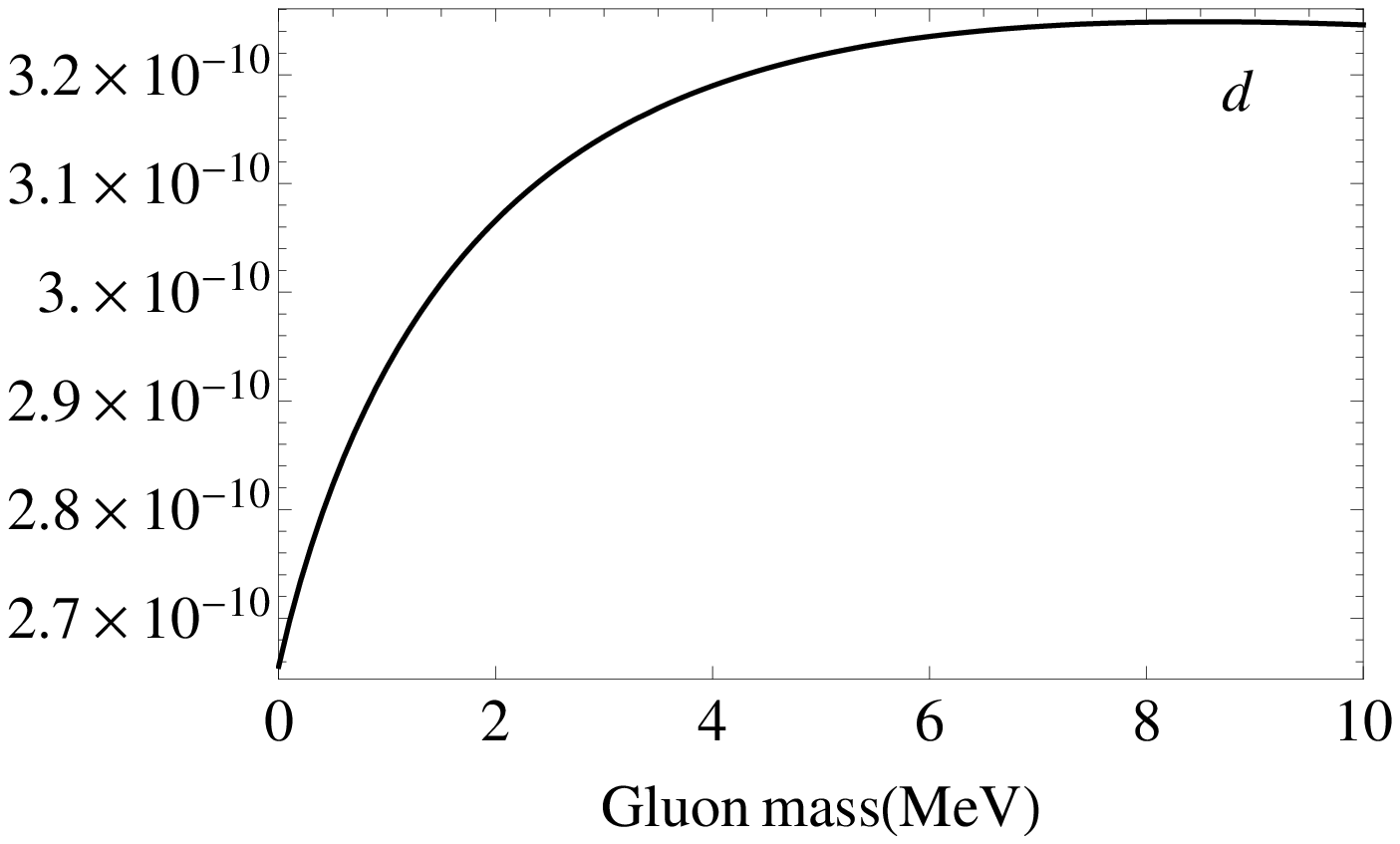}}
\hspace{0.02\textwidth}
\subfigure[]{\label{fig.mz_s}
\includegraphics[width=0.3\textwidth]{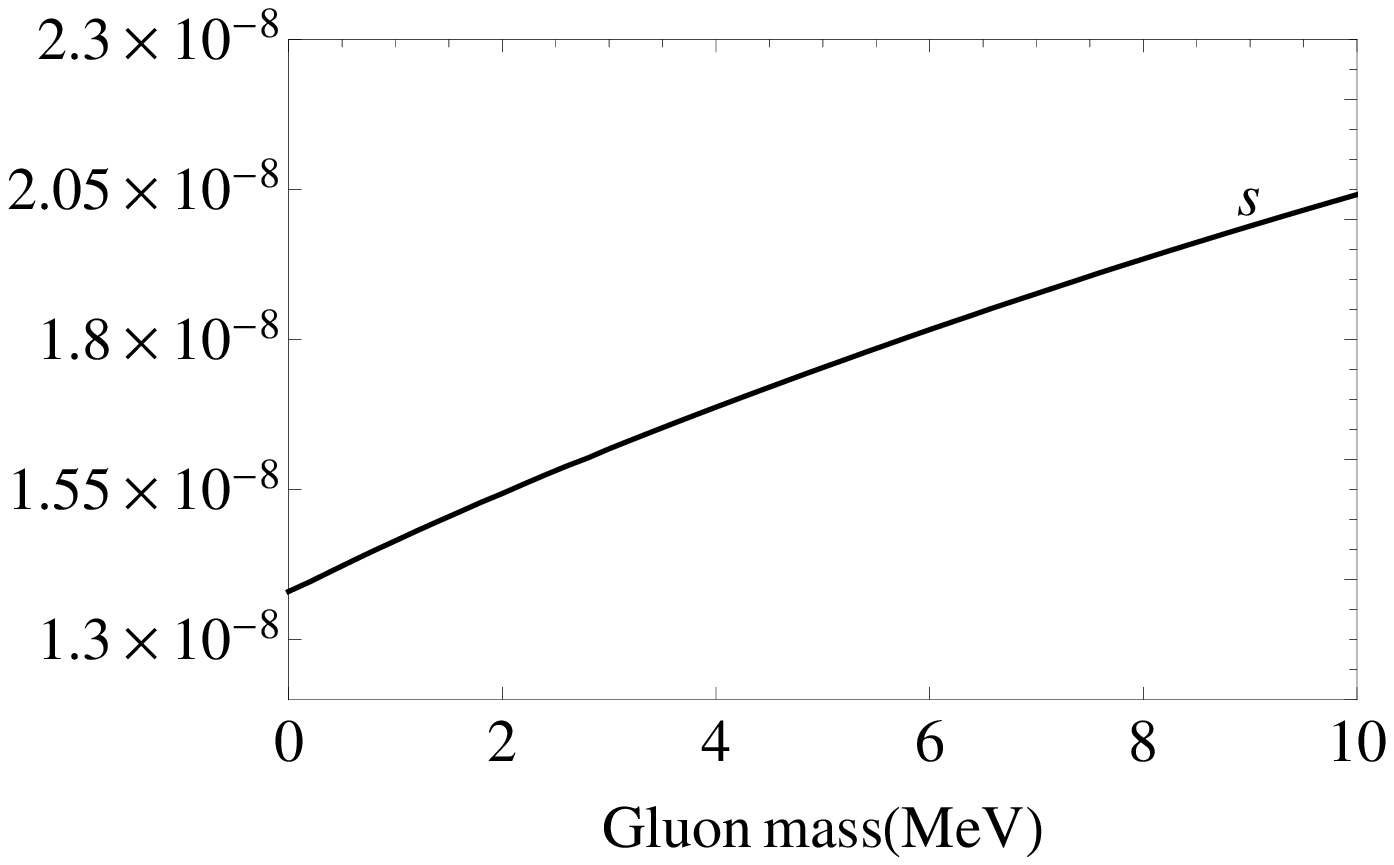}}
\\
\subfigure[]{\label{fig.mz_c}
\includegraphics[width=0.3\textwidth]{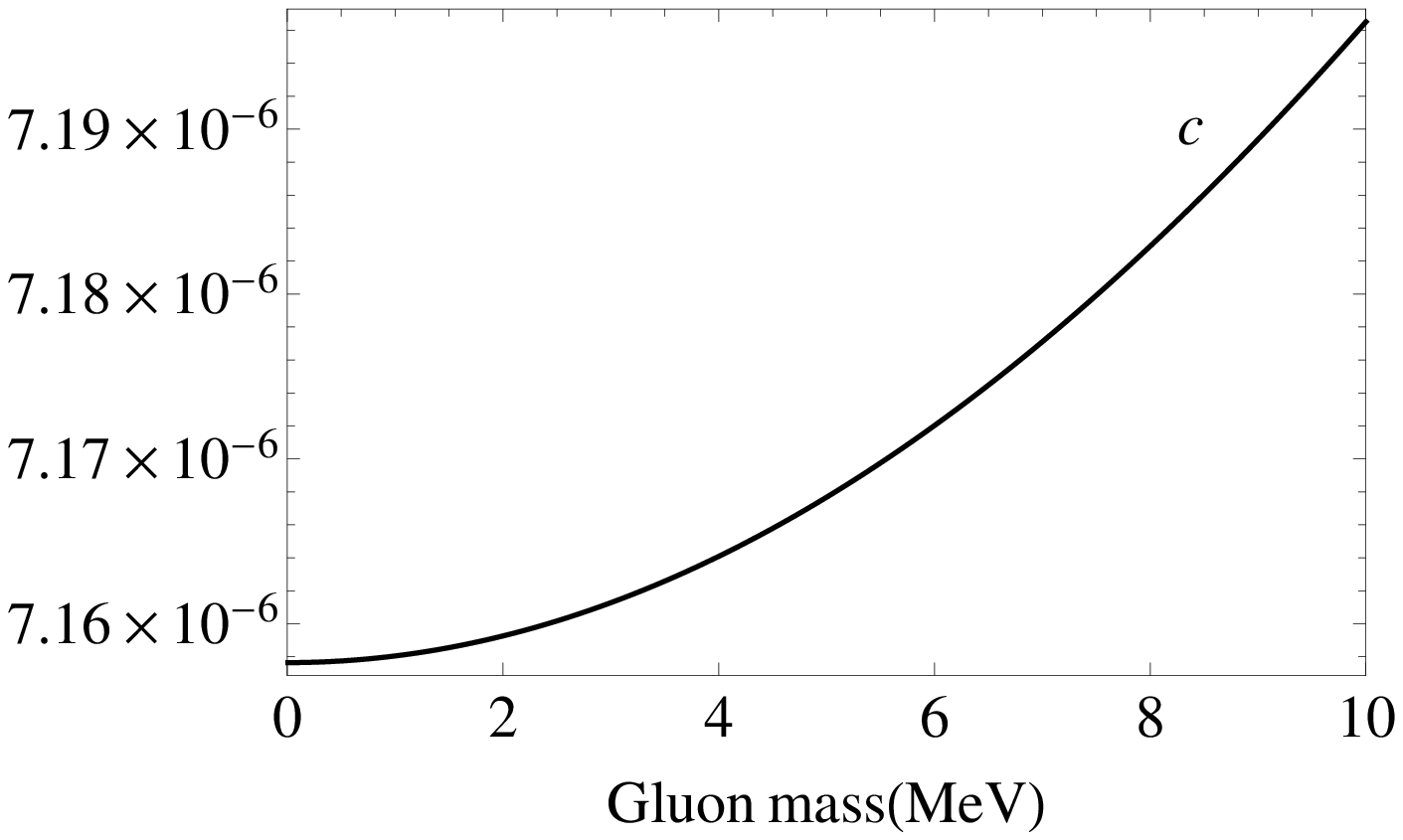}}
\hspace{0.02\textwidth}
\subfigure[]{\label{fig.mz_t}
\includegraphics[width=0.3\textwidth]{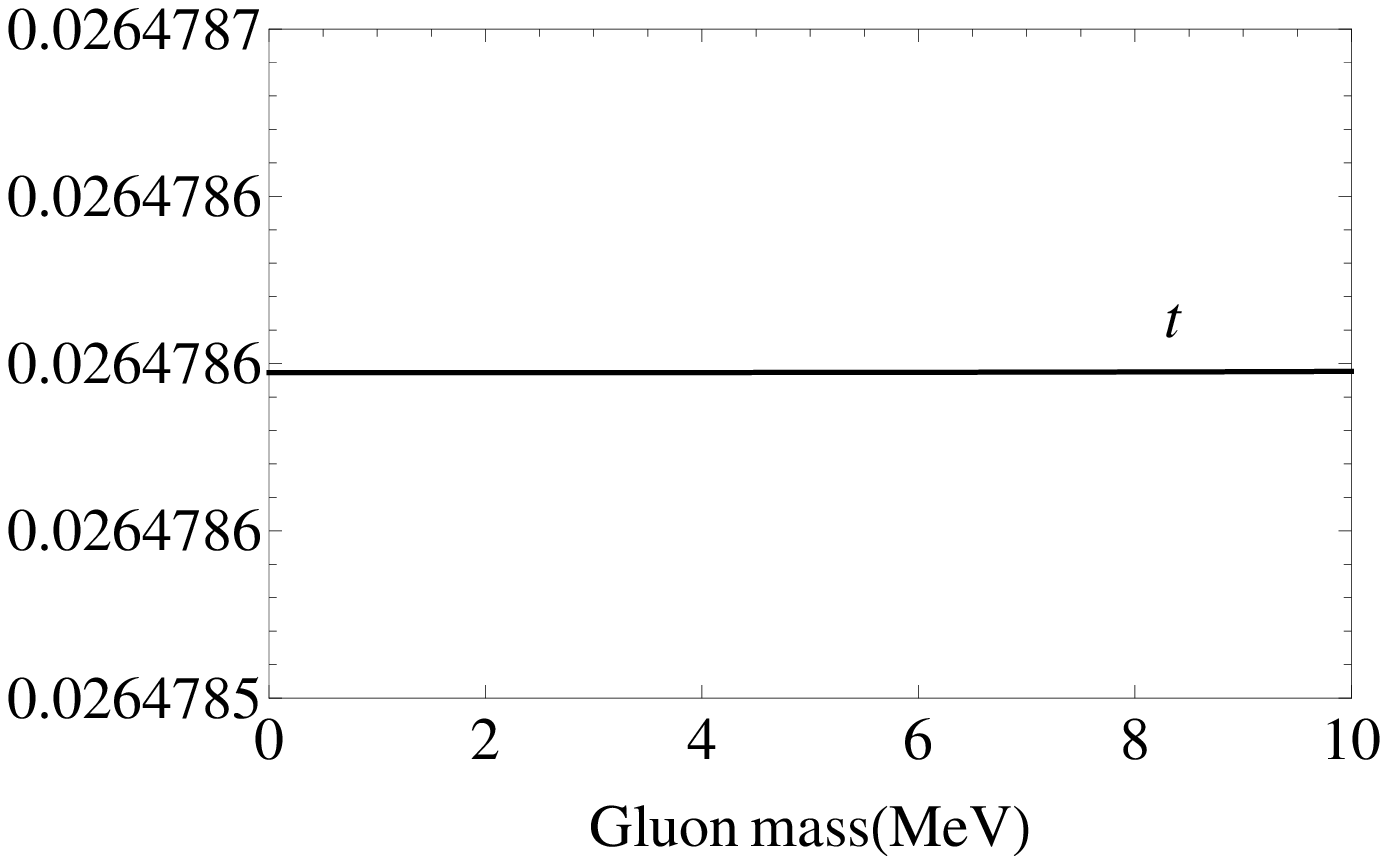}}
\hspace{0.02\textwidth}
\subfigure[]{\label{fig.mz_b}
\includegraphics[width=0.3\textwidth]{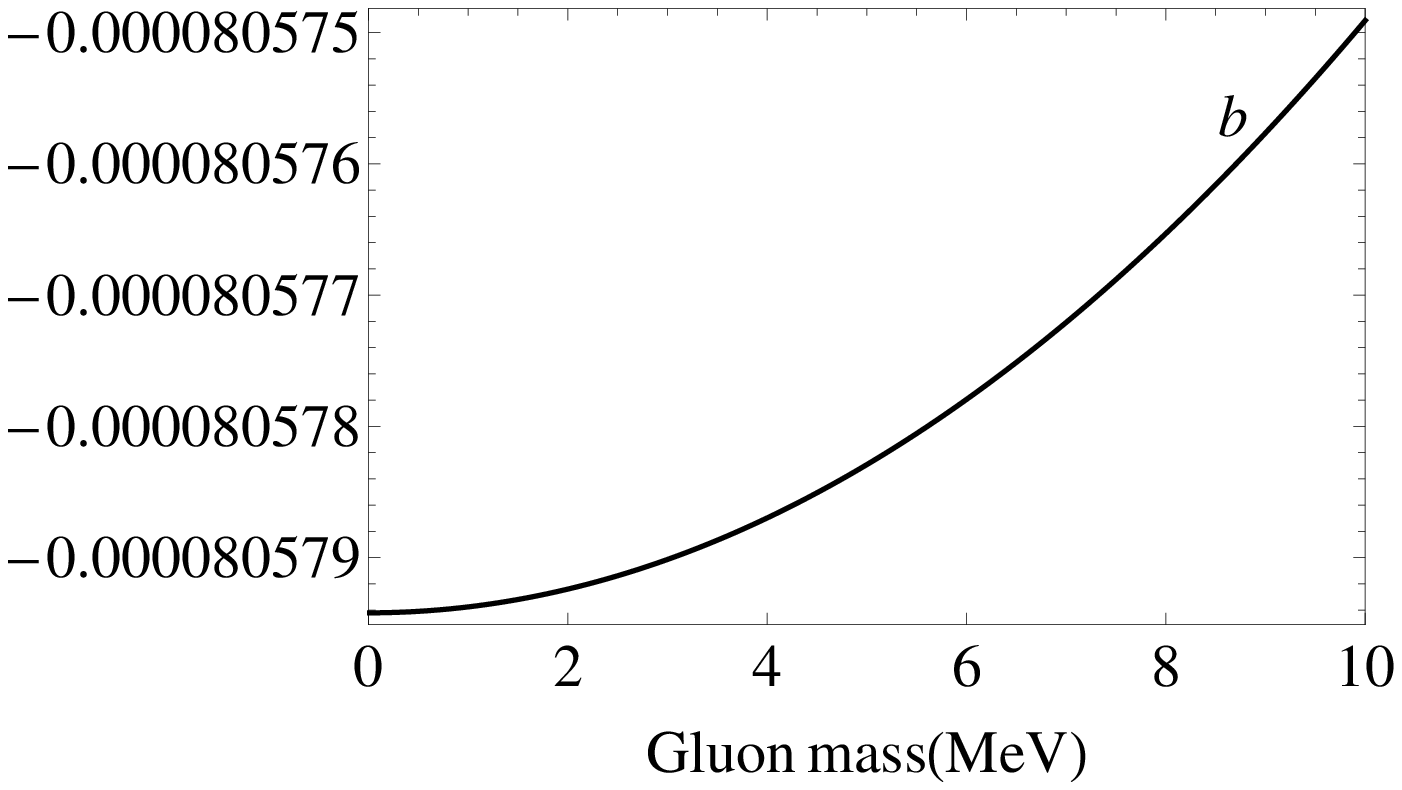}}
\caption{Anomalous CMDM of quarks at $M_Z$\,: dependence 
on gluon mass} 
\label{fig.total_mz}
\end{figure}

In Table~\ref{table:weak-MZ} we have collected the electroweak 
contributions to the form factor for the different quarks at 
$q^2 = -M_Z^2\,.$ The values of $G_F\,, m_i\,,$ and $\alpha$
were taken from~\cite{Beringer:1900zz}. The zeroes in this 
table represent numbers which are smaller by at least a factor 
of $10^{-3}$ than the smallest number in the same row.
For ease of comparison 
with the zero momentum case, we have given the values of the 
dimensionless quantity $4m F_{2}\,.$ We note that unlike for 
$q^2 = 0\,,$ this quantity differs by orders of magnitude 
between different quarks, clearly depending on the mass of 
the quark. This difference exists for the strong contributions 
as well, as we can see by adding to this the output of 
Eq.s~(\ref{F1aM_Z^2}) and~(\ref{F1bM_Z^2}). We have plotted 
the total for each quark in Fig.~\ref{fig.total_mz}\,; 
clearly it is not meaningful to plot all of them in the same 
graph.

%

\section{Discussion of results}\label{disc}
The anomalous chromomagnetic dipole moment of quarks 
may be defined analogously to the anomalous magnetic dipole 
moment. However, as we have seen in this paper, this analogy 
is not perfect. The non-Abelian nature of Yang-Mills theory
leads to an additional diagram which diverges for vanishing 
gluon mass. If the gluon has a small mass, the 
divergence goes away; it is possible to set limits on the 
mass by using experimentally observed limits on 
$\Delta\kappa\,.$

In~\cite{Martinez:2001qs}, 
$\Delta\kappa$ for the top quark
was defined and calculated at zero momentum exchange 
(on-shell gluons), and there the bounds obtained were 
$|\Delta\kappa|\leqslant 0.45$ 
from Tevatron experiments, and a more stringent bound of 
$-0.03 \leqslant\Delta\kappa\leqslant 0.01$ from 
$b\to s\gamma$ transitions measured by the CLEO collaboration.
But $\Delta\kappa$ is infrared divergent if the gluons are
massless and that the more stringent bound is not satisfied
by $\Delta\kappa$ for any gluon mass up to 10 MeV. The 
less stringent bound appears to be satisfied as long as 
the gluon mass is more than ${\cal O}$(0.1 MeV)\,. However,
it is inappropriate to use perturbation 
theory to calculate the anomalous moments at $q^2 = 0\,,$ 
and thus the relevant quantities should be calculated at 
some other value of $q^2$\,.

We have therefore calculated the 
relevant form factor at the $Z$-mass, i.e. at 
$q^2 = -M_Z^2\,.$ The results are now very different; 
there is no divergence for vanishing gluon mass. 
The dependence on gluon mass is most pronounced for 
the light quarks $u,d$ and $s\,,$ varying by 10 to 
15\% over a range of 0-10 MeV for the gluon mass. 

Discussions about the anomalous chromomagnetic dipole 
moment in the literature have focused only on the top 
quark because it is larger in magnitude than for the 
other quarks. We have corrected the existing results
for the top quark at $q^2 = 0$ and also given results
at the $Z$ mass. It is also seen from our results that
the anomalous chromomagnetic dipole moments of the 
other heavy quarks at may have measurable values at 
$q^2 = -M_Z^2$ as well. Future 
experiments at the LHC may be able to impose more 
precise bounds on the anomalous chromomagnetic 
dipole moments of different quarks, thus putting 
more stringent bounds on the mass of gluons. 

\begin{acknowledgments}
We thank C.~D.~Roberts for making us aware of~\cite{Chang:2010hb}
and related work.
\end{acknowledgments}

\end{document}